\documentclass[prd,twocolumn,preprintnumbers]{revtex4-1}

\usepackage[usenames,dvipsnames,svgnames]{xcolor}
\definecolor{redd}{rgb}{0.8, 0.1,0.2}
\definecolor{navy}{rgb}{0.05, 0.23,0.75}
\usepackage[colorlinks]{hyperref}
\hypersetup{
     colorlinks   = true,
     citecolor    = navy,
	linkcolor = redd,
	urlcolor=navy,
	anchorcolor=blue
}
\usepackage{bbm}
\usepackage{amsfonts}
\usepackage{amsmath}
\usepackage{mathrsfs}

\usepackage{amsmath,amssymb,amsbsy,bm}

\usepackage{epsfig}
\usepackage{graphicx}              
\usepackage{url}
\usepackage{hyperref}
\usepackage{float}
\usepackage{pstricks}
\usepackage{color}
\usepackage{multirow}
\usepackage{lipsum}
\usepackage{enumitem}
\usepackage{ctable}
\newcolumntype{L}{>{\centering\arraybackslash}m{1.5cm}}

\usepackage{enumitem}

\newcommand{\be}{\begin{equation}}
\newcommand{\ee}{\end{equation}}
\newcommand{\bea}{\begin{eqnarray}}
\newcommand{\eea}{\end{eqnarray}}
\newcommand{\beq}{\begin{eqnarray}}
\newcommand{\eeq}{\end{eqnarray}}

\newcommand{\msol}{M_\odot} 

\newcommand{\jcap}{JCAP}
\newcommand{\apjl}{ApJ Lett.}
\newcommand{\aap}{AAP}
\newcommand{\mnras}{Mon. Not. Roy. Astr. Soc.}

\newcommand{\ourtitle}{The Subhalo Mass Function and Ultralight Bosonic Dark Matter}
\newcommand{\twiddle}{{\raise.17ex\hbox{$\scriptstyle\sim$}}}

\newcommand\numberthis{\addtocounter{equation}{1}\tag{\theequation}}

\begin{document}
\preprint{MIT-CTP/5173}
\title{\ourtitle}
\author{Katelin Schutz}
\affiliation{Center for Theoretical Physics, Massachusetts Institute of Technology, Cambridge, MA 02139}

\begin{abstract}
\noindent 
Warm dark matter has recently become increasingly constrained by observational inferences about the low-mass end of the subhalo mass function, which would be suppressed by dark matter free streaming in the early Universe. In this work, we point out that a constraint can be placed on ultralight bosonic dark matter (often referred to as ``fuzzy dark matter'') based on similar considerations. Recent limits on warm dark matter from strong gravitational lensing of quasars and from fluctuations in stellar streams separately translate to a lower limit of $\sim 2.1 \times 10^{-21}$~eV on the mass of an ultralight boson comprising all dark matter. These limits are complementary to constraints on ultralight dark matter from the Lyman-$\alpha$ forest and are subject to a completely different set of assumptions and systematic uncertainties. Taken together, these probes strongly suggest that dark matter with a mass $\sim 10^{-22}$~eV is not a viable way to reconcile differences between cold dark matter simulations and observations of structure on small scales. 
\end{abstract}
\maketitle
\noindent

\section{Introduction}
To date, dark matter (DM) has only been detected via its gravitational influence on visible matter. Gravitational probes may yet prove to be the optimal way to understand the properties of DM if it only interacts substantially with visible matter through gravitational interactions. DM candidates with unique clustering signatures would be especially ripe for exploration in this manner. For instance, if DM decouples from the thermal bath of the early Universe while relativistic, the growth of cosmological structure will be affected. This ``warm dark matter'' (WDM) thermal history has been extensively studied to determine its effect on linear and non-linear cosmology. In this scenario, lowering the WDM mass leads to a longer free-streaming length because WDM moves too quickly to cluster for a longer period of time. Thus, lowering the WDM mass leads to a predicted suppression of structure on progressively larger and larger scales where the effects become more readily apparent in comparison to the predictions of cold DM (CDM). 

Limits on WDM have been set in the quasilinear regime using the Lyman-$\alpha$ forest as a tracer of the density field at high redshifts. WDM masses $m_\chi \lesssim 4$--5~keV are excluded by these measurements, with the exact limit depending on the modeling assumptions about the thermal history of the intergalactic medium (IGM)~\cite{Baur:2015jsy,Yeche:2017upn,Irsic:2017ixq,Baur:2017stq}. Recently, limits on WDM have become even tighter based on observational inferences about the low-mass end of the subhalo mass function (SHMF), which would be suppressed in a WDM cosmology. For instance, recent analyses of the strong gravitational lensing of quadruple-image quasars imply the existence of low-mass subhalos that are abundant enough to exclude WDM masses below $m_\chi \lesssim 5$--6~keV~\cite{2019MNRAS.tmp.2780H,Gilman:2019nap}. Additionally, a population of low-mass subhalos has recently been suggested as the origin of fluctuations in the linear density of stellar streams. When combined in a joint analysis with classical Milky Way satellite counts, the fluctuations in stellar streams exclude WDM masses below $m_\chi < 6.3$~keV~\cite{Banik:2019smi}. Aside from constraining this particular DM scenario, these results constitute measurements of the shape of the SHMF. Accordingly, the results can be interpreted in the context of any theory that predicts suppression of the SHMF.

One such theory, often referred to as ``fuzzy dark matter'' (FDM), posits that DM is an ultralight boson that is so low in mass that its de Broglie wavelength is manifest on $\sim$kiloparsec scales in galactic DM halos. For the relevant range of speeds in halos like the Milky Way, $v\sim 10^{-3}$, FDM must have a mass near $10^{-22}$~eV to have such a large de Broglie wavelength (often the convention is to write the mass of this boson in units of $10^{-22}$~eV, which we denote as $m_{22}$ in this work). Ultralight bosons, particularly axions, are compelling DM candidates in the context of string theory and their discovery could shed light on physics at extremely high energies and address the lack of observed $CP$ violation in quantum chromodynamics (see e.g. Refs.~\cite{witten1984some,svrcek2006axions,Arvanitaki:2009fg,acharya2010m,Cicoli:2012sz}). The large de Broglie wavelength of this DM candidate may reduce the abundance and central density of dwarf galaxies and subhalos as compared to CDM, which has been invoked (for instance, by Refs.~\cite{Hu:2000ke,Hui:2016ltb}) as a potential explanation for the discrepancy between observations and CDM-only simulations of these systems~\cite{Bullock:2017xww}. We collectively refer to these discrepancies in subhalo density and abundance as small-scale structure issues. Note however that the difference between observations and CDM-only simulations is plausibly explained by baryonic physics and may not pose a challenge to the CDM paradigm (see for instance Ref.~\cite{2016ApJ...827L..23W}). Also note that FDM may not be compatible with the observed properties of dwarf galaxy density profiles~\cite{Deng:2018jjz,Safarzadeh:2019sre}. Nevertheless, the possibility that DM is an ultralight boson merits consideration and a number of ideas have been put forth to constrain the unique signatures of FDM using galactic-scale observations~\cite{Marsh:2011bf,2015JCAP...12..025M,2015arXiv150300799R,Bar-Or:2018pxz,Amorisco:2018dcn,Li:2018kyk,Emami:2018rxq,Bar:2018acw,Church:2018sro,DeMartino:2018zkx,Grin:2019mub,Davies:2019wgi,Lancaster:2019mde,Mocz:2019emo}.

An analogy can be made between WDM and FDM cosmologies because FDM exhibits a similar suppression of density perturbations below a characteristic scale: in a FDM universe, the de Broglie scale is effectively a Jeans scale below which perturbations cannot grow due to the uncertainty principle. The Lyman-$\alpha$ forest has already been used to exclude $m_{22} \lesssim 20$, again with the exact number depending on modeling of the IGM~\cite{Kobayashi:2017jcf,Armengaud:2017nkf,Irsic:2017yje,Nori:2018pka}. As is the case for WDM, inferences about the SHMF may improve upon --- and independently corroborate --- the Lyman-$\alpha$ forest constraints on FDM. The details of how fluctuations are suppressed are not exactly the same for WDM and FDM, and therefore the predictions for the SHMF will not necessarily have a one-to-one mapping between the two theories. However, in this work we show that a conservative limit on FDM of $m_{22} \sim 21$ can be set based on recent SHMF WDM constraints from stellar streams and from gravitational lensing. In particular, FDM with $m_{22} \sim 21$ predicts a suppression of the SHMF at low subhalo masses that is equal to or stronger than the WDM scenarios that are excluded by stellar streams and quasar lensing.

Taken at face value, the Lyman-$\alpha$ constraints on $m_{22} \lesssim 20$ already exclude the possibility that FDM is responsible for solving any small-scale structure issues, which would require $m_{22}\sim 1$. However, independent constraints are particularly helpful in this case due to uncertainties regarding the astrophysical modeling of the IGM. Some have called the strength of the Lyman-$\alpha$ constraints into question (see for instance, Refs.~\cite{Hui:2016ltb,Garzilli:2019qki}), arguing that the key observable effect of FDM, \emph{i.e.} the suppression of the flux power spectrum on small scales, could potentially be mimicked by spatial fluctuations in the ionizing background radiation, large instantaneous temperature inhomogeneities in the IGM due to patchy reionization, or integrated IGM thermal histories that lead to nontrivial gas pressure effects. Since the path toward solidifying and improving Lyman-$\alpha$ constraints is bound to be full of nuances and theoretical challenges, there is considerable strength in having a diversity of different probes disfavoring $m_{22} \sim 1$ with different systematics and underlying assumptions. Unlike the case of the Lyman-$\alpha$ forest, thermal properties of the IGM do not affect the bound on $m_{22}$ from stellar streams or lensing, and these limits all derive from different astrophysical environments, redshifts, and clustering regimes (\emph{i.e.} quasilinear versus collapsed subhalos). Limits on FDM specifically from the Milky Way SHMF are also a very direct way to test whether FDM can be responsible for solving small-scale structure issues in the local Universe. It would be difficult to argue that FDM could explain the fluctuations in the GD-1 stellar stream while still creating a ``missing satellites problem'' in the Milky Way halo. In light of corroborating evidence from three distinct astrophysical systems (the Lyman-$\alpha$ forest, stellar streams, and lensed quasars) that $m_{22}$ must be greater than $\sim 20$, it is highly unlikely that FDM with $m_{22}\sim 1$ could be responsible for any differences between observation and CDM-only simulations of small-scale structure.

In Sec.~\ref{sec:wdm}, we give a brief overview of the observations that lead to an inference of the SHMF and describe how WDM was modeled and constrained in Refs.~\cite{Gilman:2019nap,Banik:2019smi}. In Sec.~\ref{sec:fdm}, we review the predicted shape of the SHMF in FDM based on the semi-analytic modeling of Refs.~\cite{Du:2016zcv,Du:2018wxl,Du:2018zrg} and outline our procedure for setting a limit on FDM based on WDM results. In Sec.~\ref{sec:results} we present our FDM bound and discuss the implications for other observable consequences of FDM. 

\section{Warm Dark Matter Subhalo Mass Functions} \label{sec:wdm}
WDM suppresses the growth of structure for modes that are shorter than the DM free-streaming length. Typically, WDM becomes non-relativistic in the radiation-dominated epoch, where the growth of perturbations in CDM would occur logarithmically in the scale factor $a$ via the M\'esz\'aros effect~\cite{1974A&A....37..225M}. After matter-radiation equality where perturbations grow linearly in $a$, WDM can still stream freely (although it is nonrelativistic, it is still moving too quickly to cluster) until the typical WDM speed is sufficiently Hubble damped. In this situation, the WDM transfer function can be computed with a Boltzmann code (for instance \texttt{CAMB}~\cite{Lewis:1999bs}) and is well described by the fitting form~\cite{bode2001halo,viel2005constraining,Schneider:2011yu}
\beq T_\text{WDM} \equiv \left(\frac{P_L^\text{WDM}}{P_L^\text{CDM}}\right)^{1/2} = \left[ 1 + (\lambda_\text{fs}^
\text{eff} k)^{2 \nu})\right]^{-5/\nu}\eeq
where $P_L$ is the linear matter power spectrum, $\nu = 1.12$, and the effective free-streaming length is \beq\lambda_\text{fs}^\text{eff}  = 0.07 \left(\frac{m_\chi}{1~\text{keV}}\right)^{-1.11}\hspace{-0.1cm} \left(\frac{\Omega_\text{WDM}}{0.25 }\right)^{0.11}\hspace{-0.15cm}  \left(\frac{h}{0.7}\right)^{0.22}\hspace{-0.15cm}  \text{Mpc,} \quad \quad \eeq
where $\Omega_\text{WDM}$ is the fraction of the critical density of the Universe comprised of WDM and where $h$ is defined so that the Hubble parameter is $H_0 = 100 \, h$~km/s/Mpc. A scale often introduced to parameterize the spectrum of WDM perturbations is the half-mode length scale, $\lambda_\text{hm}$, which corresponds to the scale where the amplitude of the WDM transfer function is reduced by half compared to CDM. From the transfer function, \beq \lambda_\text{hm} = 2 \pi \lambda_\text{fs}^\text{eff} (2^{\nu/5} - 1)^{-1/2 \nu} \approx 13.93  \lambda_\text{fs}^\text{eff} . \eeq
A related quantity used to parameterize the SHMF is the half-mode mass, which is the average mass contained within a mode of spatial size $\lambda_\text{hm}$, \begin{align*} &M_\text{hm} = \frac{4\pi}{3} \left(\frac{\lambda_\text{hm}}{2}\right)^3 \bar{\rho}_m \numberthis\\&= 1.65 \times 10^{10}  \left(\frac{m_\chi}{1~\text{keV}}\right)^{-3.33}  \hspace{-0.1cm} \left(\frac{\Omega_\text{WDM}}{0.25 }\right)^{1.33}\hspace{-0.1cm}  \left(\frac{h}{0.7}\right)^{2.66}\hspace{-0.1cm}  \msol ,\end{align*} 
where $\bar{\rho}_m$ is the mean background matter density. The half-mode mass enters into how the SHMF is parameterized when setting a limit on WDM from observables, although as we explain in the following subsections the authors of Refs.~\cite{Gilman:2019nap,Banik:2019smi} model WDM slightly differently, partly owing to the different demands of their respective analyses. Most notably, the two limits assume slightly different SHMF shapes given fixed $M_\text{hm}$ and slightly different concentrations and density profiles for WDM subhalos.
 
\subsection{Stellar Streams}
A constraint on WDM has recently been set through the analysis of the distribution of stars in Milky Way stellar streams~\cite{Banik:2019smi}. These streams are formed when a bound system (like a globular cluster or dwarf galaxy) is tidally disrupted and stars with different orbital energies are smeared along the direction of the stream. Because of their highly elongated spatial extent, these streams are dynamically cold (i.e. have low velocity dispersion) due to Liouville's theorem. The motion of stars in a stream is easily perturbed, for instance by a passing DM subhalo, and perturbations can manifest as gaps and spurs in the stream. Because these streams are so dynamically cold, these features can persist over gigayear timescales. Ref.~\cite{Bonaca:2018fek} used the significant gaps in the GD-1 stream to infer the existence of a perturber with a mass between $10^5 - 10^8 \msol$ that could not be accounted for by the orbits of known globular clusters, dwarf galaxies, or molecular clouds. Ref.~\cite{Banik:2019cza} generalized this idea by looking at the one-dimensional power spectrum of the stellar density along the stream rather than looking for individual gaps. Structure on $\sim10$~degree scales was found in the GD-1 stream, implying the existence of $\sim 3 \times 10^7 \msol$ subhalos. 

We note that Ref.~\cite{2020ApJ...891..161I} called the interpretation of Refs.~\cite{Banik:2019smi,Banik:2019cza} into question, claiming that the density fluctuations appear to be fairly periodic and can be reproduced by an $N$-body simulation with epicyclic motion in a smooth potential, avoiding the need for subhalos to explain the observation. However, epicyclic overdensities in stellar streams would be caused by episodic periods of tidal stripping~\cite{kupper2010tidal,kupper2012more}; this effect was already studied in Refs.~\cite{sanders2016dynamics,bovy2017linear} where epicyclic contributions to the stream density power spectrum were found to be highly subdominant in comparison to ones sourced by passing substructure because different episodes of stripping quickly mix together due to velocity dispersion in the stream. The stream in the $N$-body simulation of Ref.~\cite{2020ApJ...891..161I} was evolved for 2~Gyr; thus, to match the length of GD-1 the progenitor mass was chosen to be $3\times 10^4 \msol$. Consequently, the simulated stream was thicker than what is observed in GD-1. By comparison, other simulation work has found that to match all the observed properties of GD-1, including the width as well as the length, the stream must be older and have a lower progenitor mass~\cite{webb2019searching}. For the shorter dynamical time assumed for the simulated stream in Ref.~\cite{2020ApJ...891..161I}, stripped stars have less time to mix which accounts for the relative importance of epicyclic overdensities. In contrast, the simulations of Ref.~\cite{webb2019searching} (which do not include passing DM substructure) only have density fluctuations present near the disrupting progenitor, which is why Refs.~\cite{Banik:2019smi,Banik:2019cza} do not include the parts of the stream near the progenitor in their analysis. In summary, it seems that the limits of Refs.~\cite{Banik:2019smi,Banik:2019cza} are unlikely to be affected by the results of Ref.~\cite{2020ApJ...891..161I} due to the different assumptions in the respective analyses about the properties of GD-1; however, we acknowledge that further study of the GD-1 stream will refine our understanding of the implications for WDM. We emphasize that the SHMF measurement obtained with gravitational lensing (discussed in the next Subsection) stands completely independently of any of these nuances with stellar streams. For the remainder of this Subsection, we describe the results of Refs.~\cite{Banik:2019smi,Banik:2019cza} without further discussion of these caveats.

By analyzing the GD-1 and Palomar~5 streams and including the counts of classical Milky Way satellites, WDM below $m_\chi <6.3$~keV was excluded based on the inferred SHMF, shown in Fig.~\ref{stream}. In this exclusion, a uniform prior on the logarithm of $m_\chi$ was adopted in the range between $1-50$~keV. The WDM SHMF was parameterized as \beq \left( \frac{dN }{d \ln M}\right)_\text{WDM}^\text{stream} = \left( 1 + \gamma \frac{M_\text{hm}}{M}\right)^{-\beta} \left( \frac{dN }{d \ln M}\right)_\text{CDM} \label{gammaparam} \eeq
where the best fit in the Aquarius simulations~\cite{Springel:2008cc} is $\gamma = 2.7$ and $\beta=0.99$~\cite{Lovell:2013ola} and where the superscript label denotes that this was the WDM SHMF parameterization used in the stellar stream limit on WDM. The SHMF for CDM in this work was taken to be a power-law fit to the subhalos in the Aquarius simulations~\cite{Springel:2008cc}, 
 \beq \hspace{-0.2cm} \left( \frac{dN }{d M}\right)_\text{CDM} \hspace{-0.2cm}=  3.26\times 10^{-5} \left(\frac{M}{2.57 \times 10^7 \msol}\right)^{-1.9}\hspace{-0.1cm} \msol^{-1}.\quad  \eeq
 The subhalos are modeled as having a radial distribution in the Milky Way that follows an Einasto profile~\cite{Springel:2008cc}. In addition, for subhalos lighter than $10^9 \msol$, the normalization of the SHMF (for both WDM and CDM) is allowed to float in order to account for disruption by the baryonic content of the inner Milky Way. Tidal disruption of subhalos appears in simulations to be independent of subhalo mass~\cite{garrison2017not,graus2018through}, and is therefore not expected to change the \emph{shape} of the SHMF in this low-mass regime. The rising SHMF at low subhalo masses, shown in Fig.~\ref{stream}, constrains the shape of the SHMF and excludes WDM masses that would cause a large departure from the predictions of CDM for subhalo masses $M \sim 3\times 10^7\msol$.

For convenience, the subhalos in Refs.~\cite{Banik:2019smi, Banik:2019cza} were modeled as Plummer spheres to have the same profile as other potential perturbers of the stream, like molecular clouds and globular clusters. The Plummer profiles had a scale radius of $r_s = 1.62 (M/ 10^8 \msol)^{0.5}$~kpc~\cite{2016MNRAS.463..102E}, which yields the same one-dimensional density power spectrum on large $\sim10$~degree scales as a Hernquist profile with a similar concentration~\cite{Bovy:2016irg}. The fiducial mass-concentration relation of Ref.~\cite{Bovy:2016irg} was obtained with fits to the Via Lactea II simulation~\cite{diemand2008clumps}, and varying this relation by changing the scale radius by a factor of 2.5 was found to have a negligible impact above $\sim$degree scales in stellar streams. The inference about the low-mass end of the SHMF from stellar streams, which primarily comes from the observed structure on large scales, is therefore robust to differences in the subhalo density profile. For instance, it is not expected that assuming a Navarro-Frenk-White (NFW) profile (or its truncated version) for DM subhalos (as is done in the analysis of strong lensing as described in the next Subsection) would meaningfully change the observed effect of passing subhalos on stellar streams, given roughly similar mass-concentration relations. It is therefore also appropriate to apply the resulting measurement of the SHMF to FDM subhalos with roughly similar mass-concentration relations as the WDM case. 

\subsection{Quasar Lensing}
Another approach for constraining the low-mass end of the SHMF involves quasars that are quadruply imaged due to strong gravitational lensing by foreground galaxies. Based on the positions of the four images and their flux ratios, which depend on the second derivative of the lensing potential near the images, the presence of subhalos down to a mass of $\sim 10^7 \msol$ can be inferred with existing data. In this Subsection we focus the discussion on the recent analysis of Ref.~\cite{Gilman:2019nap} which studied the effect of WDM substructure on the lensing of eight quasars with narrow-line emission~\cite{2019MNRAS.tmp.3214N}. We note that a similar bound on WDM was independently set in Ref.~\cite{2019MNRAS.tmp.2780H} which studied the effect of substructure on the lensing of seven different radio-loud quasars using a different choice of physical priors and modeling assumptions.

Ref.~\cite{Gilman:2019nap} found an upper limit on the half-mode mass of $6.3\times 10^7\msol$, corresponding to a WDM mass of 5.2~keV. In this limit, a uniform prior was adopted in the logarithm of the half-mode mass, $\log_{10}(M_\text{hm}/\msol) \in [4.8, 10]$. The WDM SHMF was parameterized as 
\beq \left( \frac{dN }{d \ln M}\right)_\text{WDM}^\text{lensing} = \left( 1 +  \frac{M_\text{hm}}{M}\right)^{-\beta} \left( \frac{dN }{d \ln M}\right)_\text{CDM} \label{nogamma} \eeq
 where the superscript label denotes the form of the WDM SHMF for the limit from lensing, and with the best fit of $\beta=1.3$ providing a slightly worse fit to the same simulations than the parametrization including $\gamma$ (see Eq.~\eqref{gammaparam}) that was used in the stellar stream analysis~\cite{Lovell:2013ola}. The CDM SHMF used in this case comes from the publicly available \texttt{galacticus} code~\cite{Benson:2010kx}, which is particularly useful in accounting for parent halos with different masses at different redshifts. The results from \texttt{galacticus} yield a very similar CDM SHMF as the Aquarius simulations, with a power law slope for $dN / dM$ between $-1.85$ and $-1.95$ and a normalization that increases with redshift. The spatial distribution of subhalos was modeled as following the density profile of the main lens galaxy outside a tidal radius that was taken to be half the scale radius. Inside the tidal radius, the spatial distribution of subhalos was taken to be uniform. As in the analysis of stellar streams, the normalization of the SHMF was allowed to float to account for tidal disruption of subhalos by the main lens galaxy. 
 
 The line-of-sight halos in the lensing analysis were modeled as having NFW density profiles, while subhalos of the main lens halo were modeled as having tidally truncated NFW density profiles~\cite{baltz2009analytic}, with the truncation radius determined by the position inside the host halo~\cite{cyr2016dark}. In the WDM scenario, the concentration of a halos at fixed mass is reduced compared to the CDM case because of the delayed onset of structure formation. The functional form of the suppressed mass-concentration relation was taken to be \beq \frac{c_\text{WDM}}{c_\text{CDM}} = (1+z)^{0.026 z-0.04} \left(1+ 60 \frac{M_\text{hm}}{M}\right)^{-0.17}\label{lenscon} \eeq
 where the CDM mass-concentration relation of Ref.~\cite{diemer2019accurate} was used with a scatter of 0.1~dex~\cite{dutton2014cold}. The limit from strong lensing is sensitive to the central density and concentration of perturbing halos because the observable effect of subhalos on quadruply imaged quasars comes from nonlinear combinations of second derivatives of the lensing potential near the image. Less concentrated halos with lower central densities will have a less pronounced lensing signature; in the next Section, we argue that low-mass FDM subhalos likely have a smaller gravitational footprint at fixed halo mass (i.e. a lower central density and concentration) compared to the WDM subhalos considered in the lensing analysis, and it is therefore conservative to apply lensing limits on the SHMF for WDM subhalos to FDM ones. 
 
 \section{Fuzzy Dark Matter Subhalo Mass Function}\label{sec:fdm}
Unlike for the case of WDM, the SHMF has not been fully studied via direct simulation in the case of FDM, as it is computationally challenging due to the dynamical range required to fully solve the Schr\"odinger-Poisson equation in a large enough simulated box. Semi-analytic techniques pose a compelling alternative for determining the SHMF, and such techniques have already been validated for the case of WDM (see e.g.~\cite{Benson:2012su,Pullen:2014gna}). To study the FDM SHMF, Ref.~\cite{Du:2016zcv} used \texttt{galacticus}, a modular semi-analytic code for galaxy formation~\cite{Benson:2010kx}. As an input to \texttt{galacticus}, merger trees were constructed using the algorithm of Ref.~\cite{cole2000hierarchical}. The distribution of progenitor masses for the merger tree was determined using the extended Press-Schecter formalism (\emph{i.e.} numerically solving the excursion set problem), including modifications to the usual extended Press-Schecter formalism to account for the behavior of FDM. The FDM transfer functions, as computed with \texttt{AxionCAMB}~\cite{Hlozek:2014lca}, are suppressed at small scales relative to CDM due to the large de Broglie wavelength of FDM. For cosmologies with a suppressed transfer function on small scales, like WDM, using a sharp-$k$ (Fourier space tophat) window function in the extended Press-Schecter formalism provides the best match to simulations~\cite{Schneider:2014rda}, so in Ref.~\cite{Du:2016zcv} a sharp-$k$ filter is used to solve the excursion set problem. Finally, Ref.~\cite{Du:2016zcv} uses a scale-dependent excursion set barrier, as is appropriate for FDM since small-scale modes must be denser to collapse as compared to CDM in order to counteract effective quantum pressure~\cite{Marsh:2013ywa,Du:2016zcv}. A numerical prescription for the tidal disruption of FDM soliton cores inside of halos is also included in determining the SHMF, which leads to further suppression of the SHMF on the low-mass end~\cite{Du:2018zrg}. The resulting SHMF of this whole procedure is well described by the fitting form \beq \left( \frac{dN }{d \ln M}\right)_\text{FDM} = f_1(M) +  f_2(M) \left( \frac{dN }{d \ln M}\right)_\text{CDM} \label{fdmform}\eeq
with \beq f_1(M) = \beta \exp \left[ -\left(\ln \frac{M}{M_1 \times 10^8 \msol}\right)^2 /\sigma\right]\eeq
\beq f_2(M) = \left[ 1 + \left( \frac{M}{M_2 \times 10^8 \msol}\right)^{-\alpha_1}\right]^{-10/\alpha_1}.\eeq 
Including the effects of tidal stripping on the FDM core as simulated in Ref.~\cite{Du:2018zrg}, the parameters entering into the fitting form are \begin{align*} &\alpha_1 = 0.72 \quad M_1/m_{22}^{-1.5} = 4.7  \quad M_2/m_{22}^{-1.6} = 2.0 \\ &\beta/m_{22}^{1.5} = 0.014 \quad \sigma = 1.4. \numberthis \end{align*} This fitting form agrees well with the procedure outlined above up to FDM masses of $m_{22} = 50$. 
 
The FDM subhalos of Ref.~\cite{Du:2016zcv} were modeled as having NFW density profiles with an embedded soliton core and a modified concentration parameter. The subhalo mass-concentration relation for FDM is reduced compared to CDM (in analogy to WDM) because of the delayed onset of structure formation relative to CDM at a given subhalo mass. In the absence of detailed FDM simulations to directly determine typical FDM concentrations, the WDM concentration mass relation of Ref.~\cite{Schneider:2011yu} was adopted to describe FDM, 
 \beq \frac{c_\text{FDM}}{c_\text{CDM}} =  \left(1+ 15 \frac{M_\text{hm}}{M}\right)^{-0.3}, \label{cmassfdm}\eeq
 where $M_\text{hm}$ is defined analogously for FDM as it is for WDM and where the mass-concentration relation for CDM was taken from Ref.~\cite{gao2008redshift}. This is a slightly different mass-concentration relation than the one assumed in the galaxy lensing analysis (see Eq.~\eqref{lenscon}), but the two agree within $\mathcal{O}(10\%)$ near and above the half-mode mass and the concentration used for the lensing analysis is always higher than the one adopted for the FDM analogy. The direct analogy to WDM here is likely conservative, as FDM has additional, non-gravitational behavior (i.e. due to the uncertainty principle) that may lower the FDM concentration relative to WDM for a fixed subhalo mass. In concert with the central core of the FDM subhalo density profile (as opposed to the cusp of a NFW profile, which would have a higher central density than low-mass FDM subhalos~\cite{Veltmaat:2018dfz}), the reduced concentration may reduce the gravitational footprint (i.e. the observed effect on lensing) of FDM subhalos relative to their WDM counterparts for fixed subhalo mass. Ref.~\cite{Du:2016zcv} also considered the effect of changing the FDM mass-concentration relation, finding that even with substantial changes to this relation (i.e. changing the power law by a factor of 2 in Eq.~\eqref{cmassfdm}) the SHMF is not meaningfully affected. This justifies the conservative application of this SHMF for FDM to the analyses of stellar streams and quasar lensing in spite of the fact that slightly different subhalo density profiles and concentrations are considered in those analyses.
 
 \begin{figure}[t!]
 \includegraphics[width=0.48\textwidth]{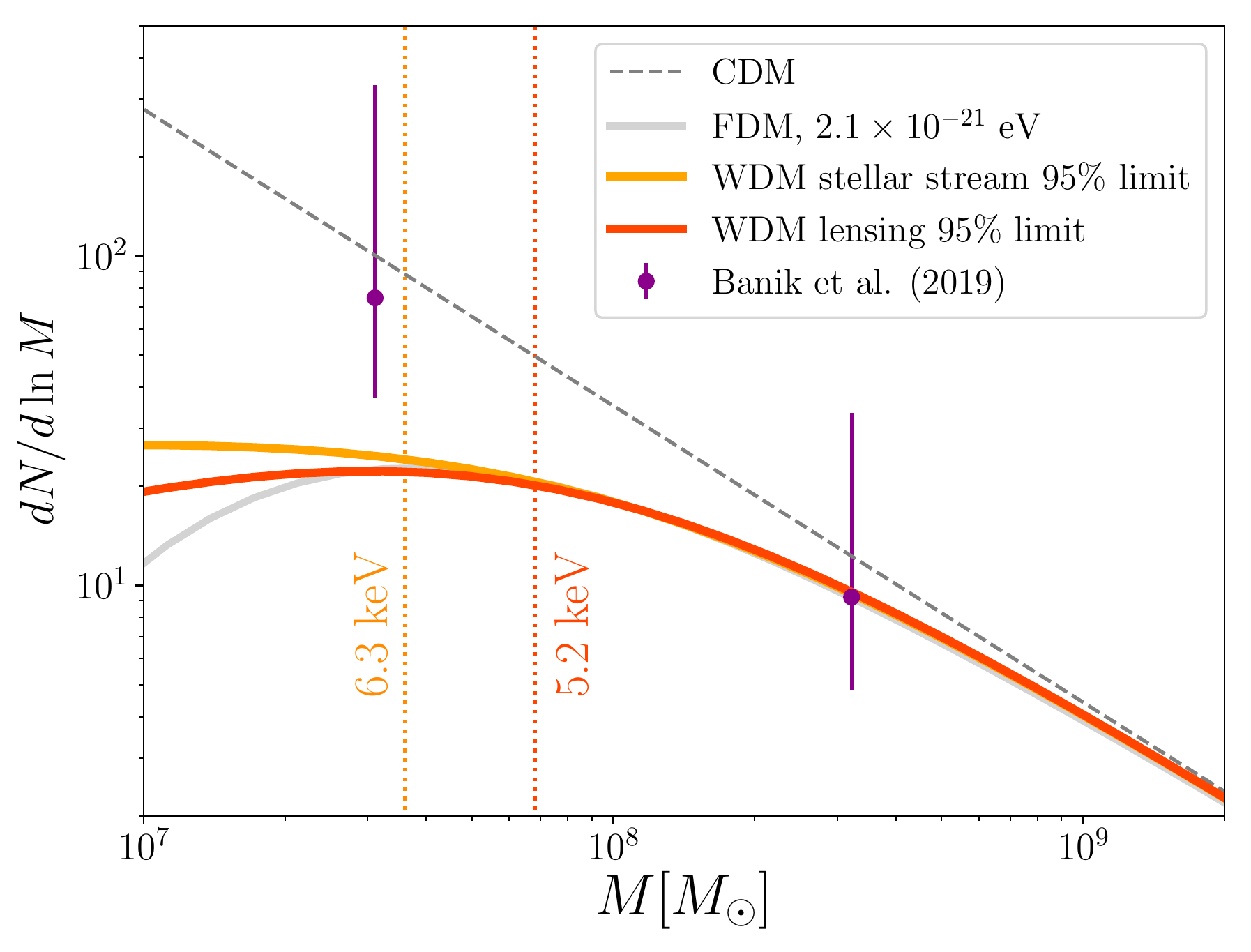}
  \caption{The SHMF for our mass limit on FDM as compared with the SHMFs for WDM that are constrained by Ref.~\cite{Banik:2019smi} from stellar streams and Ref.~\cite{Gilman:2019nap} from lensing. Vertical dotted lines show the half-mode mass $M_\text{hm}$ for the values of $m_\chi$ that are excluded in those works. The value of $m_{22}$ shown was chosen to be the maximum value of $m_{22}$ where the predicted suppression of the FDM SHMF is more dramatic than for the excluded WDM cases at all subhalo masses. In this sense, the limits on WDM can be conservatively applied to FDM. Note that all SHMFs have been normalized to match Fig.~3 of Ref.~\cite{Banik:2019smi} for subhalo masses below $\sim 10^9 \msol$, purely for the purposes of comparison of the SHMF shapes. Also note that Refs.~\cite{Banik:2019smi} and~\cite{Gilman:2019nap} model the WDM SHMF slightly differently as a function of subhalo mass, which gives slightly different SHMF shapes for fixed $m_\chi$. \label{stream}}
 \end{figure}
 
To set a limit on FDM based on existing WDM limits, the procedure we take is as follows. We normalize all SHMFs (including CDM, WDM, and FDM) in the same way, since the normalization of the SHMF for subhalo masses below $\sim 10^9 \msol$ is a free parameter in the analyses that set a limit on WDM. We are thus isolating differences in the \emph{shape} of the SHMFs at low subhalo masses for the different theories. We then vary $m_{22}$ up to the maximum value where the predicted shape of the FDM SHMF is more suppressed for all subhalo masses compared to the excluded WDM SHMFs. The limit we quote on $m_{22}$ is this maximum value where the FDM SHMF saturates the excluded WDM SHMFs. In other words, for our limit on $m_{22}$ the number of predicted subhalos in a given subhalo mass interval is always smaller for FDM than it is for excluded WDM scenarios. Increasing $m_{22}$ past this point would lead to a slight excess of FDM subhalos in some mass intervals as compared with the excluded WDM scenarios.

 \section{Results and Discussion}
 \label{sec:results}
 
\begin{figure}[t!]
 \includegraphics[width=0.48\textwidth]{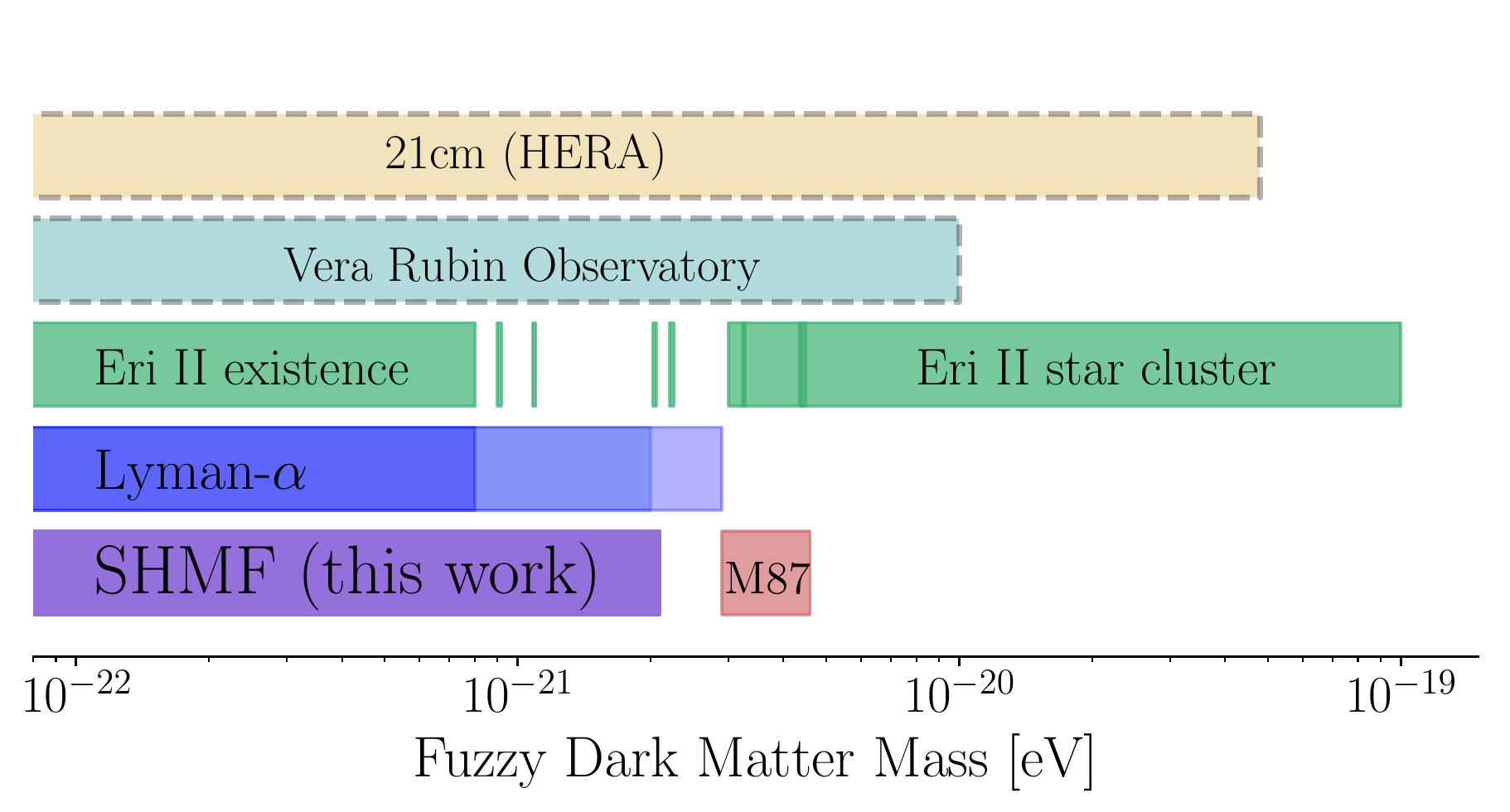}
  \caption{Summary of constraints on the FDM mass (shaded regions with no border are currently excluded) as well as projections of sensitivity (shaded regions with a dashed border can be explored in the future). The M87 bound comes from the Event Horizon Telescope measurement of the spin of M87*, the supermassive black hole in the center of M87. A scalar particle with a mass in the shaded region would cause M87* to spin down excessively due to superradiance~\cite{Davoudiasl:2019nlo}. The different shaded regions of the Lyman-$\alpha$ forest bound correspond to different analyses that make different assumptions about the IGM. The darkest region comes from matching the WDM bound of Ref.~\cite{Garzilli:2019qki} onto FDM using Eq.~(8) of Ref.~\cite{Armengaud:2017nkf}; the middle region corresponds to the bound of Refs.~\cite{Irsic:2017yje,Nori:2018pka}; the lighter region extending to slightly higher masses corresponds to Ref.~\cite{Armengaud:2017nkf}. The existence of Eridanus~II and the survival of its star cluster set limits over a range of FDM masses, including in narrow orbital resonance bands~\cite{Marsh:2018zyw}. In the future, the Vera Rubin Observatory will be able to probe even higher FDM masses up to $\sim 10^{-20}$~eV by probing the SHMF at lower subhalo masses~\cite{Drlica-Wagner:2019xan}. The observation of fluctuations in the cosmological 21~cm signal from HERA will probe even higher FDM masses~\cite{Munoz:2019hjh}. Note that all bounds in this figure assume that non-gravitational interactions are negligible. All bounds except the M87 bound assume that FDM accounts for all DM.  \label{constraint}}
 \end{figure}

The SHMFs for various DM scenarios are shown in Fig.~\ref{stream}. The SHMFs for WDM correspond to the WDM mass limits considered in this work (including the fact that the respective analyses considered slightly different WDM SHMF shapes for fixed half-mode mass, see Eqs.~\eqref{gammaparam} and~\eqref{nogamma}). Also shown is the FDM SHMF with $m_{22} = 21$, the limit determined by the procedure outlined in the previous Section (this limit of $m_{22}=21$ happens to be almost exactly the same when we compare the FDM SHMF to the WDM limits from either stellar streams or quasar lensing). Because the SHMF for $m_{22} = 21$ is even more suppressed than the WDM SHMFs that are already constrained at the 95\% level, the limit is likely to be slightly too conservative. In Fig.~\ref{constraint}, we depict this limit on the FDM mass along with others to show the full landscape of constraints.

Similar methods to the ones presented here were employed in Ref.~\cite{Marsh:2018zyw}, where the SHMF was used to set a limit on FDM based on the existence of the ultrafaint dwarf galaxy Eridanus~II. The existence of a single subhalo with a mass like Eridanus~II excludes FDM masses $m_{22} <8$. The limit in the present work is stronger because of the larger number of subhalos inferred at low subhalo masses. As shown in Fig.~\ref{constraint}, bands in $m_{22}$ are separately excluded above $m_{22} = 8$ in that work based on the survival of the star cluster in Eridanus~II, which would be dynamically heated and disrupted by fluctuations in the FDM core density. Other works have claimed limits on FDM based on the presence of Milky Way subhalos. For instance, Ref.~\cite{Nadler:2019zrb} found $M_\text{hm} <3.1\times 10^8\msol$ based on an analysis of classical and SDSS-discovered Milky Way satellites and converted this to a limit on FDM $m_{22} >29$ using the formula \beq m_{22} = 13 \times \left(\frac{\Omega_m}{0.25}\right)^{0.95} \left(\frac{h}{0.7}\right)^{1.9} \left(\frac{M_\text{hm}}{10^9 \msol }\right)^{-0.71} \label{match}\eeq based on Eq.~(8) of Ref.~\cite{Armengaud:2017nkf}. However, this formula is based on the mapping between the half-mode scales of FDM and WDM in the quasilinear regime and does not account for the full $k$-dependence of the FDM transfer function. Based on the hydrodynamical simulations and analysis of Ref.~\cite{Irsic:2017yje}, the matching of half-mode scales between FDM and WDM is not accurate for translating a Lyman-$\alpha$ forest limit on WDM to one on FDM. We also expect that in the case of the SHMF, the matching of the WDM and FDM transfer functions at the half-mode scale will not necessarily provide an accurate limit on FDM because this approach does not take into account the scale-dependent growth of structure (and scale-dependent critical collapse overdensity) in a FDM cosmology. Had we set a limit by matching half-mode scales in this work, we would have found that FDM masses below $m_{22} \lesssim 137$ are excluded. We instead quote a more conservative bound of $m_{22} \lesssim 21$ here. 
 
The use of stellar streams has previously been proposed as a means to constrain FDM via the dynamical heating that would occur through the sustained density fluctuations exhibited in FDM halos, caused by wave interference on the scale of the de Broglie length~\cite{Amorisco:2018dcn}. The signature would be a marked thickening of the stream with time due to frequent interactions with FDM clumps. The strength of this effect has a steep scaling with $m_{22}$, and at the level of the constraint from the SHMF $m_{22} \gtrsim 20$ is not expected to yield an observable heating effect. The effective mass of FDM clumps for this value of $m_{22}$ is $\sim10$s of solar masses, leading to a thickening rate of $\lesssim 0.1'/$Gyr for the GD-1 stream. Since the angular thickness of GD-1 is around $12'$~\cite{Grillmair:2006bd}, dynamical heating from ambient FDM of this mass is not expected to play an observable role in the properties of the stream. Because of the rather low effective mass of FDM clumps created by wave interference for $m_{22} \gtrsim 20$, it is also not expected that such density fluctuations will have an observable signature in the lensing of quasars.

The constraint presented here is intended to serve as a conservative bound that is set without a full re-analysis of the data presented in Refs.~\cite{Gilman:2019nap,Banik:2019cza}. The FDM mass $m_{22}$ constrained here was chosen to always give a more dramatic suppression of the SHMF compared to the WDM scenarios that were already constrained with more complete analyses. In spite of the fact that the arguments presented here are likely quite conservative, we can already independently corroborate Lyman-$\alpha$ forest limits on FDM using two additional systems that require very different modeling assumptions and that have very different systematics. Taken together, the evidence strongly disfavors the possibility that FDM could have a meaningful impact on any small-scale structure issues or that FDM with $m_{22}\sim 1$ could lead to a core in the inner Milky Way and be responsible for the dynamics of bulge stars, as proposed in Ref.~\cite{DeMartino:2018zkx}. In the near-term future, the bound on the FDM mass based on the SHMF is likely to improve (both in strength and in accuracy) with dedicated analyses of the data of Refs.~\cite{Banik:2019cza,Gilman:2019nap}. A preliminary analysis of the data from stellar streams indicates that the true limit on FDM may be closer to $m_{22} >22$ assuming logarithmic priors on $m_{22}$ between 1 and 1000~\footnote{Private communication with Jo Bovy.}. Further improvement on FDM mass limits will be possible in the coming years with the Vera Rubin Observatory, which will probe the SHMF at even lower subhalo masses through a combination of detecting fainter dwarf galaxies, tracking perturbations in more stellar streams, and increasing the sample of known strongly lensed systems~\cite{Drlica-Wagner:2019xan}. Measurements of the cosmological 21~cm signal from an instrument like the Hydrogen Epoch of Reionization Array (HERA)~\cite{DeBoer:2016tnn} will also be able to detect low-mass halos that host the first stars and galaxies, probing FDM up to a mass $m_{22} \sim 480 $~\cite{Munoz:2019hjh}. 

Further simulation work will also improve our understanding of the FDM SHMF, especially as it relates to the effect that fragmented FDM filaments (as seen for instance in Ref.~\cite{Mocz:2019emo}) could have on the SHMF. It is encouraging that after this work first appeared on arXiv, Ref.~\cite{Benito:2020avv} found a similar limit on FDM using the same observables considered here but with the SHMF fitting form of Ref.~\cite{schive2016contrasting}; this fitting form was determined by matching to simulations performed with FDM initial conditions (so the fragmentation of filaments would contribute to the SHMF in the simulation) but without taking into account the effects of quantum pressure on nonlinear collapse. The limit presented here therefore seems fairly robust to differences in the FDM SHMF determined through these different methods, but further study is certainly warranted. Baryons can also have nontrivial effects on FDM (see for instance Refs.~\cite{2018MNRAS.478.2686C,Bar:2018acw,Mocz:2019emo,Mocz:2019uyd,Veltmaat:2019hou}) and further study of these effects and whether they impact the SHMF at very low subhalo masses will make the results presented here more robust.

Other DM scenarios beyond WDM and FDM may be constrained by the measurements of the SHMF, particularly ones that predict suppression of small-scale power. Some examples are DM freeze-in through a light vector mediator~\cite{Dvorkin:2019zdi} and velocity-independent DM-baryon scattering~\cite{Nadler:2019zrb}. Furthermore, in this work we have treated FDM as lacking any substantial non-gravitational self-interaction that could affect cosmological structure (see for instance Refs.~\cite{Guth:2014hsa,Desjacques:2017fmf,Arvanitaki:2019rax}). We leave exploration of the implications of the SHMF for these scenarios to future work.

 \section*{Acknowledgements}
It is a pleasure to thank Nilanjan Banik, Andrew Benson, Simon Birrer, Jo Bovy, Xiaolong Du, Joshua Eby, Denis Erkal, Daniel Gilman, Tongyan Lin, Adrian Liu, Philip Mocz, Tracy Slatyer, and Mark Vogelsberger for useful conversations pertaining to this work, as well as Daryl Haggard and Victoria Kaspi for camaraderie while stranded at Montr\'eal-Trudeau International Airport where this work was conceived. The author acknowledges the importance of equity and inclusion in this work and is committed to advancing such principles in her scientific communities. The author also acknowledges support from a Pappalardo Fellowship from the MIT Department of Physics.
\bibliography{FDMstream}

\end{document}